\newif\ifFIG
\FIGtrue

\newif\ifINCLUDE
\INCLUDEtrue

\newif\ifINCLUDENOT
\INCLUDENOTfalse

\documentclass[a4paper,11pt]{article} 

\usepackage{graphicx}

\usepackage{fullpage}
\usepackage{latexsym}
\usepackage{amsfonts}
\usepackage{amsmath}

\usepackage{subfig} 
\usepackage{epsf}
\usepackage{epstopdf}

\usepackage{xcolor}

\newcommand{\figurespath}{.} 

\newcommand{\Rmean} 	{y} 

\newcommand{\RmeanMODEL} 	{\hat{y}} 

\newcommand{\GrandRTmean} {\overline{\RTmean}}

\newcommand{\ntrials}{N_{i}} 

\newcommand{\nchosen}{n} 

\newcommand{\nstrengths}{S} 

\newcommand{\sprime}{s'} 

\newcommand{\scale}{w} 

\newcommand{\RT}{\tau} 

\newcommand{\var}{\mathrm{var}}

\newcommand{\jlog}	{\log_{2}}

\newcommand{\bits}{\:{\rm bits}}

\newcommand{\RTmean}{RT}

\newcommand{\be}        { \begin{equation}  }
\newcommand{\ee}        { \end{equation}	}
\newcommand{\bea}       { \begin{eqnarray}  }
\newcommand{\eea}       { \end{eqnarray}    }

\begin{document}

\title{\bf Using Information Theory to Measure Psychophysical  Performance}

\author{
James V Stone \\
Psychology Department, Sheffield University, England.\\
Email: {\it j.v.stone@sheffield.ac.uk} \: \\
File: combineRTbinary2021\_v11a.tex. 
}

\maketitle

\noindent
\section*{Abstract}
Most psychophysical experiments discard half the data collected. Specifically, experiments discard reaction time data, and use binary responses  (e.g. yes/no) to measure performance. Here, Shannon's information theory is used to define Shannon competence $s'$, which depends on the mutual information between stimulus strength (e.g. luminance) and a combination of reaction times and binary responses. Mutual information is the entropy of the joint distribution of responses minus the residual entropy after a model has been fitted to these responses. Here, this model is instantiated as a proportional rate diffusion model, with the additional innovation that the full covariance structure of responses is taken into account. Results suggest information associated with reaction times is independent of (i.e. additional to) information associated with binary responses, and that reaction time and binary responses together provide substantially more than the sum of their individual contributions (i.e. they act  synergistically). Consequently, the additional information supplied by reaction times suggests that using combined reaction time and binary responses requires fewer stimulus presentations, without loss of precision in psychophysical parameters. Finally, because $s'$ takes account of both reaction time and binary responses, (and in contrast to $d'$) $s'$ is immune to speed-accuracy trade-offs, which vary  between observers and experimental designs. 


\section*{Two Sentence Summary}
%
When presented with a stimulus, the observer's binary responses (e.g. yes/no) and their associated reaction times depend on the stimulus strength (e.g. luminance). The amount of Shannon information gained by an observer 
estimated from a combination of binary responses and reaction times was substantially larger than the sum of 
information gains based on separate analyses of binary responses and reaction times, implying that  binary responses and reaction times have a synergistic effect on the information gained by an observer. 

\newpage

\section*{Technical Summary} \label{techsummary}
%
The method consists of estimating the Shannon information gained by an observer when presented with a stimulus. 
This is estimated as the mutual information between stimulus strength (e.g. luminance) and a combination of  binary responses (e.g. yes/no) and their associated reaction times. Combined responses for a single observer were used to estimate a 2x2 covariance matrix $U$, which (under Gaussian assumptions) determines the entropy $H(U)$ of the joint distribution of binary responses and their reaction times. A proportional rate diffusion (PRD) model was fitted to each observer's combined response data. The model's residual noise defines a 2x2 covariance matrix $V$, which determines the entropy $H(V)$ of the joint conditional distribution of binary responses and their reaction times. The mutual information between stimulus strength and each observer's mean responses is $I = H(U) - H(V)$ bits; further analysis was used to estimate the mutual information $I_{single}$ between stimulus strength and each observer's individual responses. Model fitting was achieved by finding parameter values that maximise $I$. 
The rate at which an observer acquired information was estimated as $R = I_{single}/ {\overline{\RTmean}_{dec}}$ bits/s, where $\overline{\RTmean}_{dec}$ is an  observer's mean decision time, which is a parameter of the fitted PRD model.  

\newpage

\section{Introduction}
%
In a typical psychophysical experiment, half the data collected from each observer is discarded. Specifically, binary responses (e.g. yes/no) are used to estimate parameters such as threshold, whereas reaction times are usually discarded. Here, we use Shannon's information theory \cite{ShannonWeaverBook,StoneInformationBook2014} 
to combine binary responses with reaction times  to define {\em Shannon competence} $s'$. This has two key advantages: 
a)  fewer stimulus presentations are required to achieve a given precision in estimates of psychophysical parameters, 
b) in contrast to conventional measures (e.g.  $d'$), Shannon competence $\sprime$ is immune to the effects of observer-specific speed-accuracy trade-offs.
 
\ifFIG
\begin{figure}[b!] 
\begin{center}
\subfloat[]{\includegraphics[width=0.5\textwidth] {\figurespath/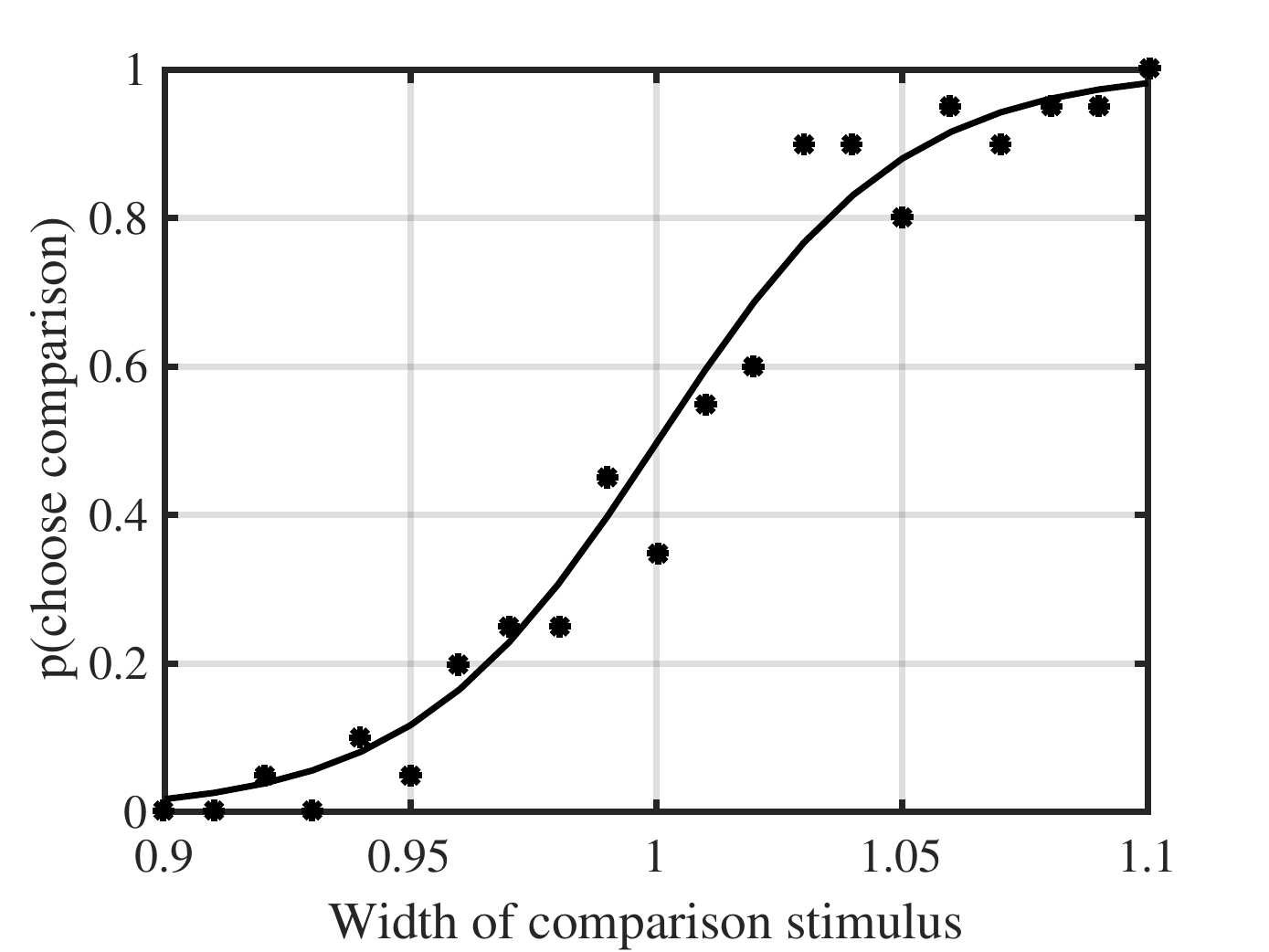} }
\subfloat[]{\includegraphics[width=0.5\textwidth] {\figurespath/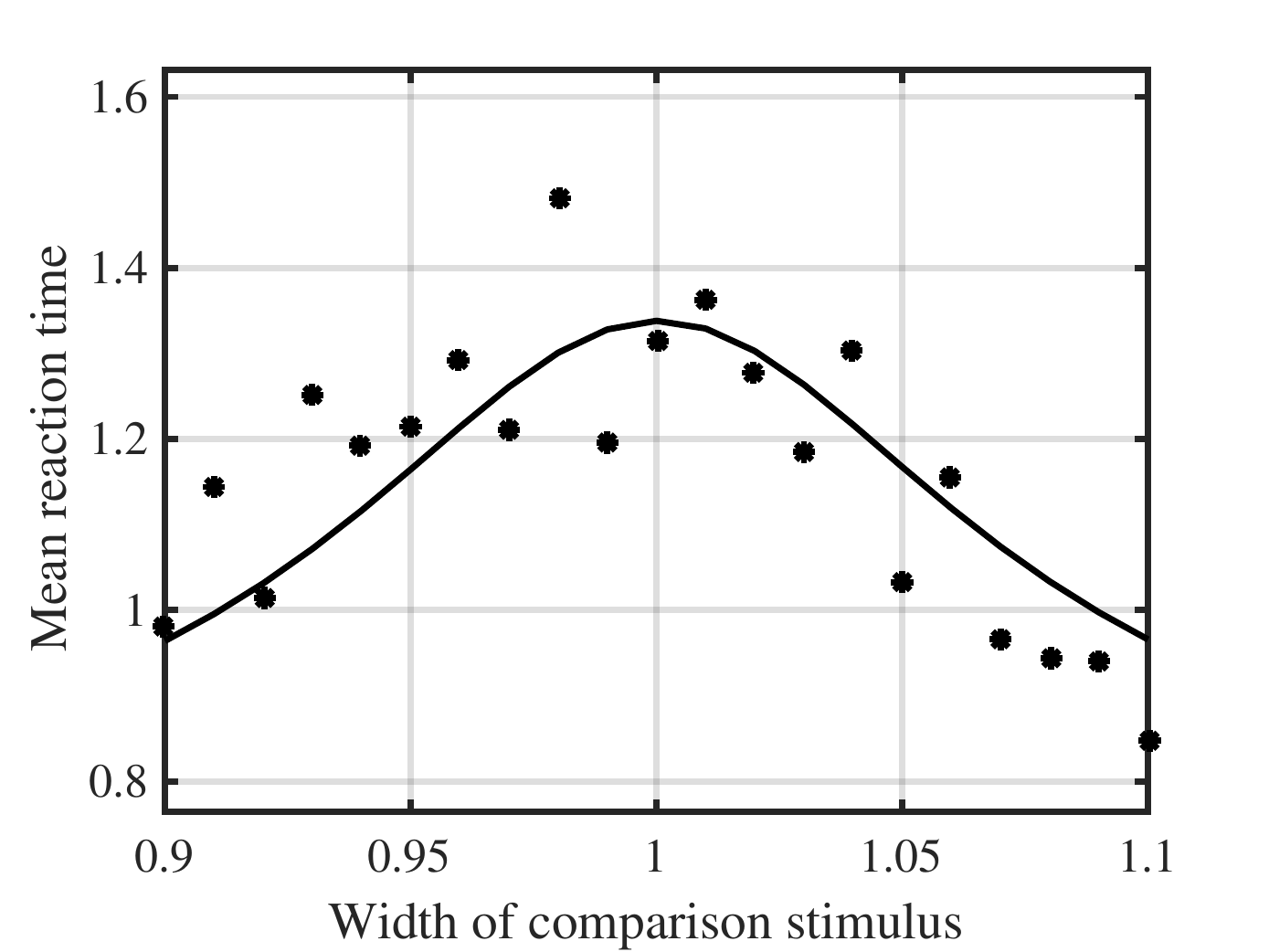} }
\caption{%
The psychometric function (a) and chronometric function (b),  from the face inversion experiment for one observer (see Figure \ref{figfaces}).
The width scaling factor $\scale$ applied to the comparison image 
is indicated on the abscissa. The stimulus strength is effectively $x=\scale-1$. 
There were 21 discrete stimulus strengths, and the stimulus at each strength was presented $\ntrials=20$ times.  
\newline
(a) 
Each dot represents the observed proportion of trials for which the observer chose the
comparison  stimulus, and the solid curve is the fitted psychometric function. 
 \newline
(b)  Each dot represents the $RT$ of a single trial for the same responses as in Figure \ref{figpsych1}a ($RT$s greater than 2 seconds are not shown). The solid curve is the fitted chronometric function. 
}
\label{figpsych1}
\end{center}
\end{figure}
\fi

\ifINCLUDE
 Here, we assume at two-alternative forced choice (2AFC) design, in which a pair of stimuli is shown to an observer, who then decides which stimulus is brighter (for example).  The {\em reference stimulus} remains constant across trials, whereas the {\em comparison stimulus} varies (e.g. in brightness) between trials. 
Each presentation of a stimulus pair defines a single trial, and the (signed) difference between stimuli defines the stimulus strength $x_{i}$. 
The stimulus strength varies over $\nstrengths$ different values, and each stimulus pair is presented $\ntrials$ times 
at each  stimulus strength $x$. The observer's response consists of a binary response $B$ and the reaction time $RT$ of that response.

If the observer chooses the comparison stimulus $\nchosen_{i}$ times at a given stimulus strength $x_{i}$ then the probability of choosing the comparison stimulus is estimated as the proportion 
\bea
	P_{i}  & = & \nchosen_{i}/\ntrials.
\eea
The probability $P_{i}$ of choosing the comparison stimulus increases as the stimulus strength comparison  increases, as shown in Figure \ref{figpsych1}a.  For completeness, the proportion of correct responses is
\bea
	{P^{c}_{i}} & = &  \frac {1} { 1+ \exp  [- |\log P_{i}/(1-P_{i})| ]  }. 
\eea

If the reaction time to $j$th presentation of the  stimulus pair with signal amplitude $x_{i}$ is $\RT_{j}$ then 
the mean reaction time at $x_{i}$ is
\bea
	\RTmean_{i} & = & \frac{1}{\ntrials} \sum_{j=1}^{\ntrials} \RT_{j}.
\eea
As the comparison stimulus strength increases, the mean reaction time $\RTmean$ increases until the stimulus strength of the comparison stimulus matches the stimulus strength of the reference stimulus, and then $\RTmean$ decreases again, as shown in Figure \ref{figpsych1}b. Consequently, the mean $RT$ increases until $x$ equals zero, and then decreases again as $x$ continues to increase.

The combined mean response of a single observer at stimulus strength $x_{i}$ can be represented with the vector variable 
\bea
	\Rmean_{i} & = & (\RTmean_{i},P_{i}).
\eea

\section{Measuring Mutual Information}
Ultimately, performance is limited by the amount of Shannon information \cite{ShannonWeaverBook} $I_{obs}(y;x)$  an observer gains when presented with   stimuli with strength $x$. Crucially, $I_{obs}(y;x)$ cannot be less than the mutual information $I(\Rmean;x)$ between stimulus strength and $\Rmean$, the combined mean binary responses and their associated mean $RT$s,
\bea
	I_{obs}(\Rmean;x) \geq I(y;x) \text{ bits}.
\eea
In other words, the observer gains an average of at least $I(x;\Rmean)$ bits of information when presented with stimulus pairs at strength $x$. We cannot measure $I_{obs}(y;x)$, but we can measure $I(x;\Rmean)$, which provides a lower bound for $I_{obs}(y;x)$. 

The mutual information can be obtained as the entropy $H(\Rmean)$ of the observer's responses minus  the residual entropy in those responses conditioned on the stimulus strength $x$,  
\bea
	I(x; y)  & = & H(\Rmean) -  H(\Rmean|x)   \text{ bits}. \label{eqMI43}
\eea
When expressed in terms of geometric areas in Figure \ref{figMI}, mutual information between stimulus strength $x$ and an observer's mean responses $y$ is
\bea
	a+b+c & = & (a+b+c+e+f+g) - (e+f+g)  \text{ bits}.
\eea

\subsection*{Evaluating  Unconditional Entropy $H(\Rmean)$}

For a given observer, the grand mean reaction time is $ \overline{\RTmean}$, and  the overall proportion of trials on which the observer chooses the comparison stimulus is  $\overline{P}$. 
The observer's grand mean response, taken over all trials and stimulus strengths,  is
\bea
	\overline{\Rmean} &= &  (\GrandRTmean, \overline{P}),
\eea
 where  $ \overline{P}$ is the grand mean value of $P_{i}$ for a given observer 
 \bea
	\overline{P} & = & \frac{1}{\nstrengths}\sum_{i=1}^{\nstrengths}  P_{i}.
\eea
and $\GrandRTmean$ is the grand mean value of $RT_{i}$ for a given observer 
 \bea
	\GrandRTmean & = & \frac{1}{\nstrengths}\sum_{i=1}^{\nstrengths}  \RTmean_{i}.
\eea
%
\ifFIG
\begin{figure}[b!] %
\begin{center}
{\includegraphics[width=0.7\textwidth] {\figurespath/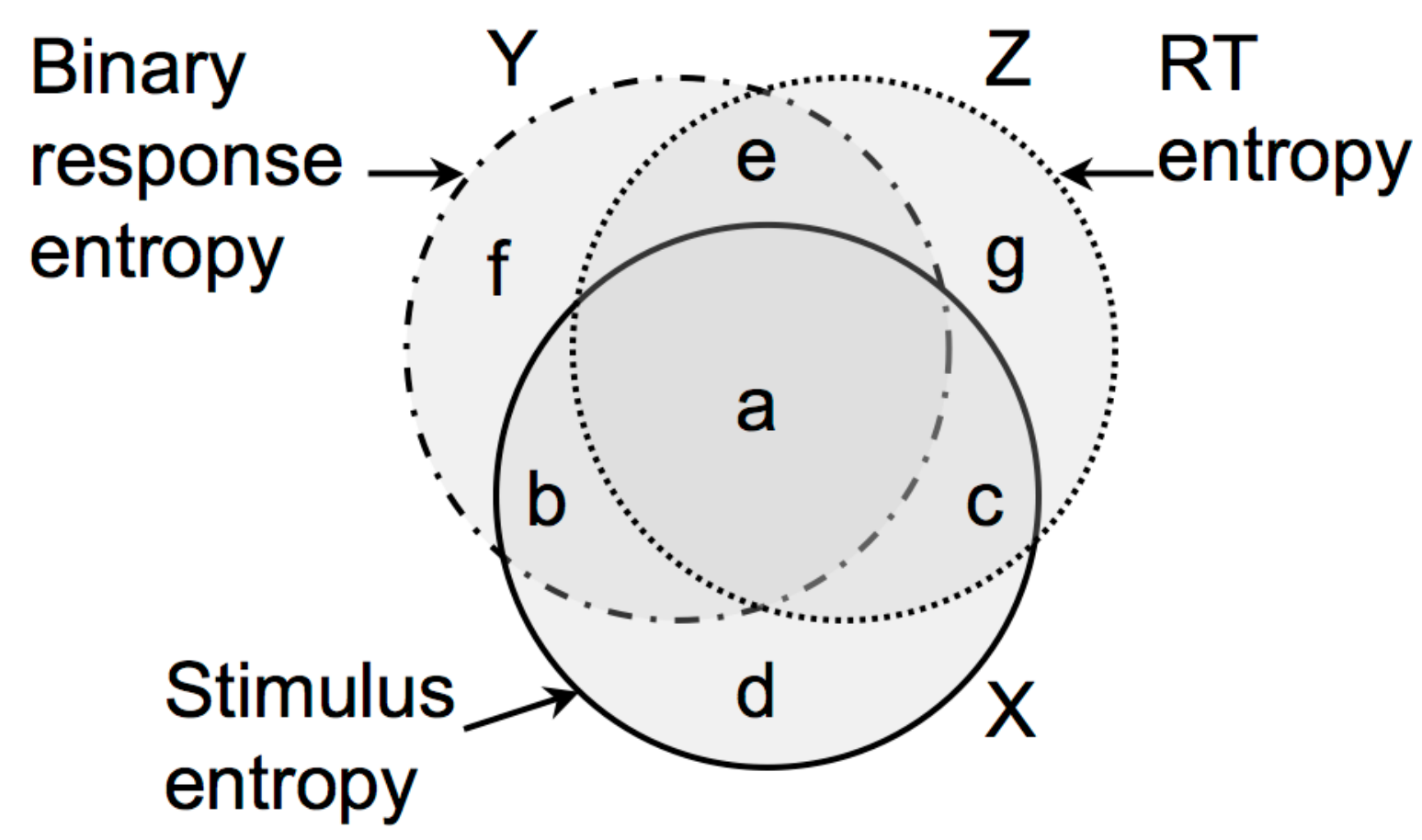} }
\caption{%
How the entropy $H({x})$ in stimulus strength $x$ is accounted for by the entropy $H(RT)$ in mean reaction time $RT$ and entropy $H(P)$ in the probability $P$ of a particular binary response. The entropies of $x, P$ and $RT$ are represented by the discs $X, Y$ and $Z$, respectively. The mutual information between $x$ and $P$ is $I(x;P)=(a+b)$, the mutual information between $x$ and $RT$ is $I(x;RT)=(a+c)$, and  the mutual information between $x$ and the combined response $RT,P$ is $I(x;RT,P)=a+b+c$.  
}
\label{figMI}
\end{center}
\end{figure}
\fi
The noise in  $RT$ at stimulus strength $x_{i}$ is
\bea
	 \eta^{\RTmean}_{i} & = & \RTmean_{i} - \GrandRTmean.
\eea
Similarly, the noise in $P_{i}$  is 
 \bea
	\eta^{P}_{i} & = & P_{i} - \overline{P}.
 \eea
The vector-valued noise in the mean responses at stimulus strength $x_{i}$ is then 
\bea
	\eta_{i}			& = & ( \eta^{\RTmean}_{i} ,\eta^{P}_{i} ).
\eea
If  the joint distribution of noise in  $\Rmean$ is Gaussian then 
the probability (density) is 
\bea 
	p(\Rmean_{i}) & = & \frac{1}{  (2 \pi)^{k/2}  |U|^{1/2}} \:  \exp \left( {-\frac{1}{2} \eta_{i} U^{-1} \eta_{i}^{T}} \right), \label{equncondpdf}
\eea
where $k=2$ is the number of variables (i.e. $\RTmean$ and $P$), as shown in Figure \ref{figpdfs}a, 
$T$ is the transpose operator, 
and 
  $|U|$ is the determinant of the the covariance matrix 
\bea
	U 
		& = &  \left( \begin{array}{cc}
				\var (\eta^{\RTmean} )		&{\rm cov}( \eta^{\RTmean}, \eta^P  ) \\
				\\
	 			 {\rm cov} (\eta^{\RTmean}, \eta^P )  	& \var (\eta^P )
			\end{array} \right),
\eea
where cov represents covariance
\bea
	{\rm cov} (\eta^{\RTmean},\eta^P ) & = & {\rm cov} (\RTmean,P ) \\
		 & = & \frac{1}{\nstrengths} \sum_{i=1}^{\nstrengths} (\RTmean_{i}-\GrandRTmean) (P_{i} - \overline{P}),
\eea
and var represents variance, 
\bea
	\var (\eta^{\RTmean} )	= \var (\RTmean )	=  \frac{1}{\nstrengths} \sum_{i=1}^{\nstrengths} (\RTmean_{i} - \GrandRTmean)^{2} \\
		\var (\eta^P ) = \var (P )	 =  \frac{1}{\nstrengths} \sum_{i=1}^{\nstrengths} (P_{i} - \overline{P} )^{2}.
\eea
The determinant $|U|$ is a generalised measure of variance, which indicates the overall spread of the distribution of $RT, P$ values, and is obtained as
\bea
	|U| & = & \var (\RTmean) \times  \var (P) ]  -  {\rm cov}( \RTmean ,P )  . \label{eqdetU}
\eea
For example, a simple rotation of axes in Figure \ref{figpdfs}a ensures that the covariance terms become zero, and then it is more obvious that  each variance term indicates the length of one axis of an ellipse, so the product of variances is proportional to the area (spread) of the ellipse.

\ifFIG
\begin{figure}[b!] 
\begin{center}
\subfloat[]{\includegraphics[width=0.5\textwidth] {\figurespath/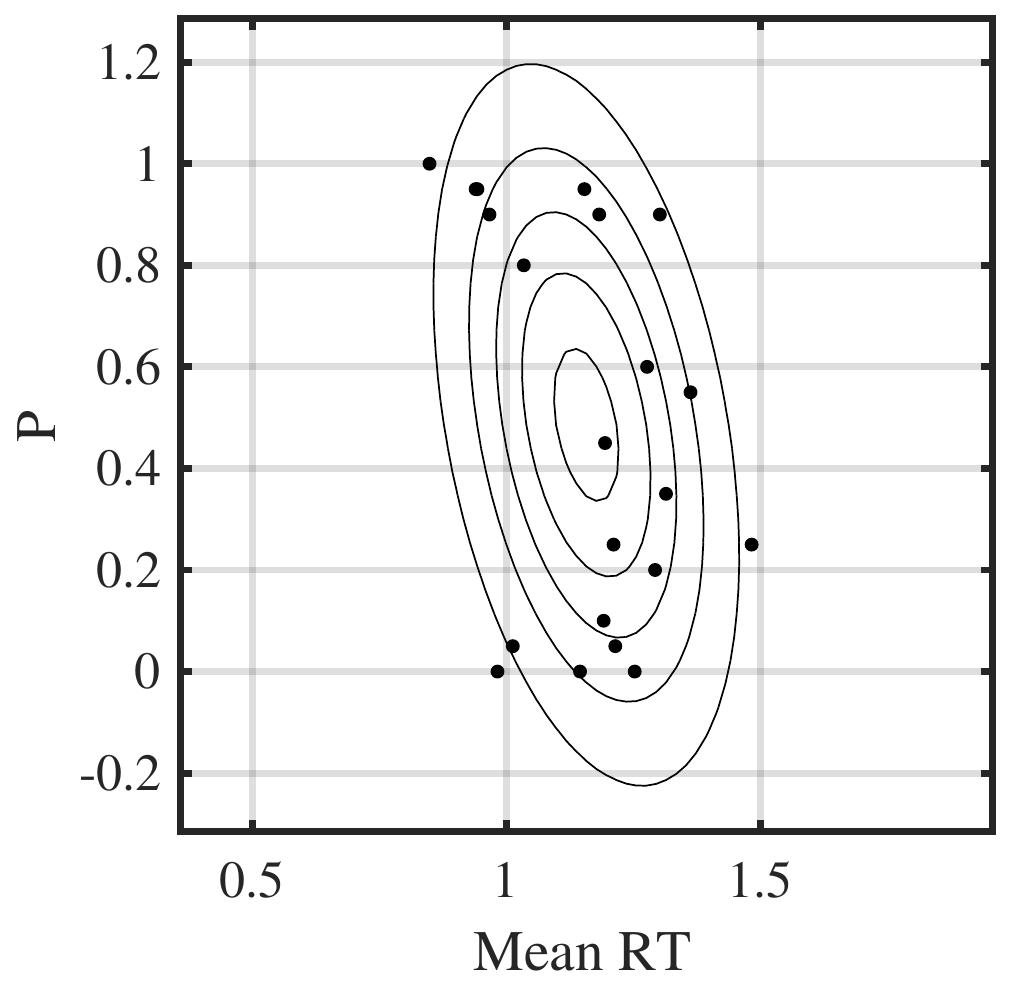} }
\subfloat[]{\includegraphics[width=0.5\textwidth] {\figurespath/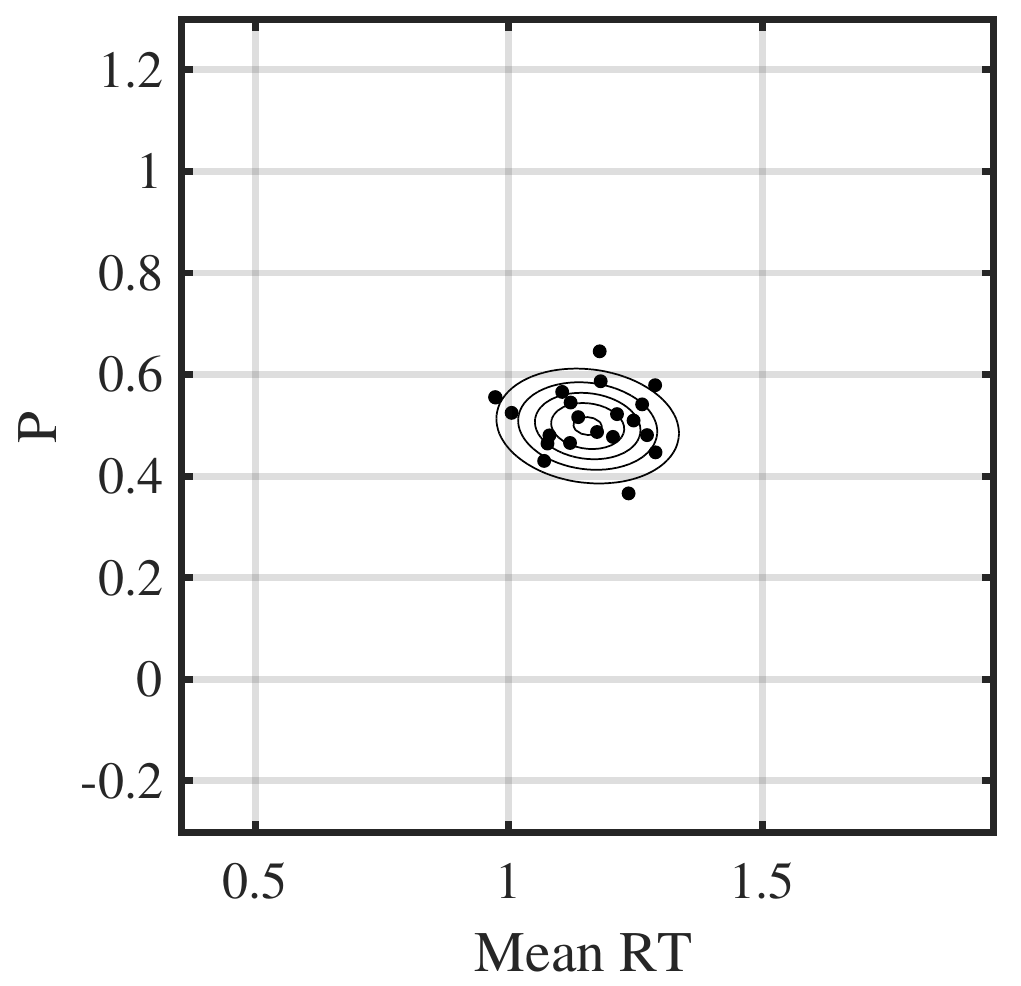} }
\caption{
Contour maps of fitted joint Gaussian distributions for a typical observer. \\
a) Joint distribution of observer's data (Equation \ref{equncondpdf}). 
Each dot represents the observer's mean response  $(\RTmean_{i},P_{i})$ to one stimulus strength $x_{i}$, 
where $P_{i}$ is the probability of choosing the comparison stimulus and $\RTmean_{i}$ is the corresponding mean reaction time at stimulus strength $x_{i}$.    \\
b) Joint distribution of residual noise, after the model has been fitted to the observer's data. 
 }
\label{figpdfs}
\end{center}
\end{figure}
\fi

Finally, the (differential) entropy of a Gaussian distribution with covariance matrix $U$ is
\bea 
	H(\Rmean)  & =&  0.5 \jlog \: (2 \pi e)^{2} | U |  \text{ bits},   \label{eqHC} 
\eea
where logarithms have base 2, which ensures that entropy is measured in bits. 

\subsection*{Evaluating Conditional Entropy $H(y|x)$}
The term $H(y|x)$ in Equation \ref{eqMI43} is the entropy of the joint distribution $p(\RTmean,P|x)$, which corresponds to the area $(e+f+g)$ in Figure \ref{figMI}. This is the average residual uncertainty in the value of $(\RTmean,P)$ for a given stimulus strength. In order to evaluate $H(y|x)$, we follow the logic of the previous section. However, instead of 
estimating noise as the difference between the observer's mean response
$\Rmean_{i}$  and that observer's grand mean, 
noise is estimated as the difference between $\Rmean_{i}$ and a model-based estimate $\hat{\Rmean}_{i}=(\hat{\RTmean}_{i}, \hat{P}_{i})$, where $\hat{\RTmean}_{i}$ is the model's estimate of the mean reaction time, and   $\hat{P}_{i}$ is the model's estimate of the proportion of trials on which  the comparison stimulus was chosen.  
Accordingly, the {\em model noise} in  the mean response  at stimulus strength $x_{i}$ is
\bea
	\varepsilon_{i} 	& = &  ( \varepsilon^{\RTmean}_{i} ,\varepsilon^P_{i} ), 
\eea
where
\bea
	 \varepsilon^{\RTmean}_{i} & = &  {\RTmean_{i} - \hat{\RTmean}_{i} }\\ 
	 \varepsilon^P_{i} & = &   P_{i} - \hat{P}_{i}.
\eea
If  the joint distribution of  noise in  $\RmeanMODEL_{i}$ is Gaussian  then 
\bea
	p(y_{i} |x_{i}) & = & \frac{1}{ (2 \pi)^{k/2} |  V |^{1/2}} \:  \exp \left( {-\frac{1}{2} \varepsilon_{i} V^{-1} \varepsilon_{i}^{T}} \right),  \label{eqmodelpdf}
\eea
where $|V|$ is the determinant of the covariance matrix 
   \bea
	V 		& = &  \left( \begin{array}{cc}
				\var ( \varepsilon^{\RTmean} ) 	 		&  {\rm cov} ( \varepsilon^{\RTmean}, \varepsilon^P ) \\
				\\
	 			  {\rm cov} ( \varepsilon^{\RTmean}, \varepsilon^P )	&  \var ( \varepsilon^{P} ) 	
			\end{array} \right).
\eea
The (differential) entropy of a Gaussian distribution with covariance matrix $V$ is
\bea 
	H(V)  & =&  0.5 \jlog \: (2 \pi e)^{2} | V |  \text{ bits},   \label{eqHCV} 
\eea
where the determinant of $V$ is 
\bea
	|V| & = & [ \var ( \varepsilon^{\RTmean} ) 	  \times   \var ( \varepsilon^{P} ) ]  -   {\rm cov} ( \varepsilon^{\RTmean}, \varepsilon^P )  . \label{eqdetV}
\eea
Finally, substituting Equations \ref{eqHC}  and \ref{eqHCV} into Equation \ref{eqMI43} yields
\bea 
	I(x; y)  	
				& = & 0.5 \jlog \: \frac{| U |}{|V|} \:\:  \text{bits},  \label{eqHUmV}
\eea
where $|U|/|V|$ is the ratio of elliptical areas in Figures \ref{figpdfs}a and \ref{figpdfs}b. 
Notice that the absolute range of reaction times and binary responses has no effect on the mutual information because this is based on the ratio variances before and after fitting a model to the data. 

\subsection*{Information Gained from Responses Versus Mean Responses}
%
So far we have derived an expression for the mutual information in the average response $y_{i}$ to a stimulus strength $x_{i}$, where this average is taken over $\ntrials$ trials. 
The average mutual information  in the observer's response in a single trial is 
\bea
	I({y_{single}} ; x) & \geq & \frac{1}{2} \jlog \left[ 1+ \frac{ 2^{2 I({x,y})}-1 }{N_{i}} \right] \bits,\label{eqSNRsingleRTy}
\eea
with equality if distributions are Gaussian 
(see Appendix). 

\subsection*{Estimating Mutual Information for RT}  
The observer entropy based only on reaction time is
\bea
	H(\eta^{RT}) & = & 0.5 \jlog 2 \pi e \, \var (\eta^{\RTmean} )	 \text{ bits}. \label{eqent1d}
\eea
The model entropy based only on reaction time is
\bea
	H(\varepsilon^{RT}) & = & 0.5 \jlog 2 \pi e \, \var (\varepsilon^{\RTmean} )	 \text{ bits}. \label{eqent1da}
\eea
 The mutual information based on reaction time is then given by
\bea
	I^{RT} & = &H(\eta^{RT}) - H(\varepsilon^{RT})    =  0.5 \jlog  \frac{ \var (\eta^{\RTmean} ) }{ \var (\varepsilon^{\RTmean} )}   \text{ bits}.  \label{eqMI1d}
\eea

\subsection*{Estimating Mutual Information for P}  
By analogy,  the mutual information based on binary responses is then given by
\bea
	I^{P} & = &H(\eta^{P}) - H(\varepsilon^{P})  =  0.5 \jlog  \frac{ \var (\eta^P ) }{ \var (\varepsilon^P )}   \text{ bits}.  \label{eqMI1dP}
\eea

\section{The CEPRD Model}  
We require a model $M$ which has a 
'response' to a stimulus strength $x_{i}$, where this response is a mean reaction time $\hat{\RTmean}$ and a probability $\hat{P}_{i}$
\bea
	\hat{y}_{i}  = (\hat{\RTmean}_{i},\hat{P}_{i}) = M(\theta,x_{i}), \label{eqmodel}
\eea
where $\theta$ is  a vector of parameters, defined below.   
The model $M$ can be instantiated as a proportional rate diffusion (PRD) model \cite{Shadlen2005}. 
Here, we use of the extended PRD (EPRD) model\cite{StoneRTInfoTheory2014}; this differs from the PRD model by making use of the probability $P$ that an observer chooses a comparison stimulus, and by  incorporating a point-of-subjective-equality parameter $x_{PSE}$.  
 The model's mean $RT$ at stimulus strength $x_{i}$ defines the chronometric function
\bea
	\hat{\RTmean}_{i} & = & \hat{\RTmean}_{dec,i} +  \hat{\RTmean}_{m} \:\:\:\: {\rm seconds}, \label{eqhataubar}
\eea
where $\hat{\RTmean}_{m}$ is the component of reaction time required for physical movement once a decision has been made, and where the mean time taken to decide which response to make is 
\bea
			 \hat{\RTmean}_{dec,i} & = &   \frac{A}{K (x_{i}-x_{PSE}) } \tanh(AK (x_{i}-x_{PSE}))  \:\:\:\: {\rm seconds}.  \label{eqRTmean}
\eea
  The  perceived stimulus strength is 
 \bea
	 x_{i}-x_{PSE}, \label{eqpse}
\eea 
where $x_{PSE}$ is the signal strength $x$ at which the reference and comparison stimuli are perceived as being the same. 
The model response probability $\hat{P}_{i}$ at stimulus strength $x_{i}$ defines the psychometric function
\bea
	\hat{P}_{i} 	
		& = & \tanh(AK  (x_{i}-x_{PSE}) )/2+1/2. \label{eqhatP}
\eea
Solving for $ \tanh(AK  (x_{i}-x_{PSE}) )$ and substituting in Equation \ref{eqRTmean},  the relation between the psychometric and chronometric functions is 
\bea
	 \hat{\RTmean}_{dec,i} & = & \frac{A}{K} \frac{2 \hat{P}_{i} -1} { (x_{i}-x_{PSE}) }   + \hat{\RTmean}_{m}  \:\:\:\: {\rm seconds},  \label{eqhatRT}
\eea
From Equation \ref{eqhatP}, the probability that an observer chooses the comparison stimulus 
depends on the product $AK$. In contrast, (from Equation \ref{eqhatRT}) decision time depends on the ratio $A/K$ and on the product $AK$.  

The model parameters for a single observer can be represented as the vector
\bea
	\theta & = & (A,K,\RTmean_{m},x_{PSE}). 
\eea
Model parameters are estimated using Equation \ref{eqmodelpdf} to define the log likelihood
\bea
	L(\theta) 	& = & \sum_{i=1}^{\nstrengths}  \jlog p( y_{i} | x_{i} ).  \label{eqMLE2}
\eea
Values of the parameters $\theta$ that maximise $L$ are found using the simplex search algorithm \cite{NUMERICAL_RECIPES_BOOK}. 
After fitting the model to an observer's data, Equation \ref{eqhataubar} provides a value of $\hat{\RTmean}_{i}$ and Equation \ref{eqhatP} provides a value  $\hat{P}_{i} $,  for a given a stimulus strength $x_{i}$. These equations were used to plot the solid curves in Figures \ref{figpsych1}a and \ref{figpsych1}b.

Note that estimating model parameters by maximising $L(\theta)$ represents an improvement on the methods reported in \cite{Shadlen2005} and \cite{StoneRTInfoTheory2014}. In those papers, model parameters were estimated by minimising the product $ \var (\varepsilon^P )  \times  \var (\varepsilon^{\RTmean} )$, which amounts to ignoring the covariance between $P$ and $RT$. To differentiate between these models, the current version is called the covariant EPRD (CEPRD) model. 


\ifFIG
\begin{figure}[b] %
\begin{center}
{\includegraphics[width=0.4\textwidth] {\figurespath/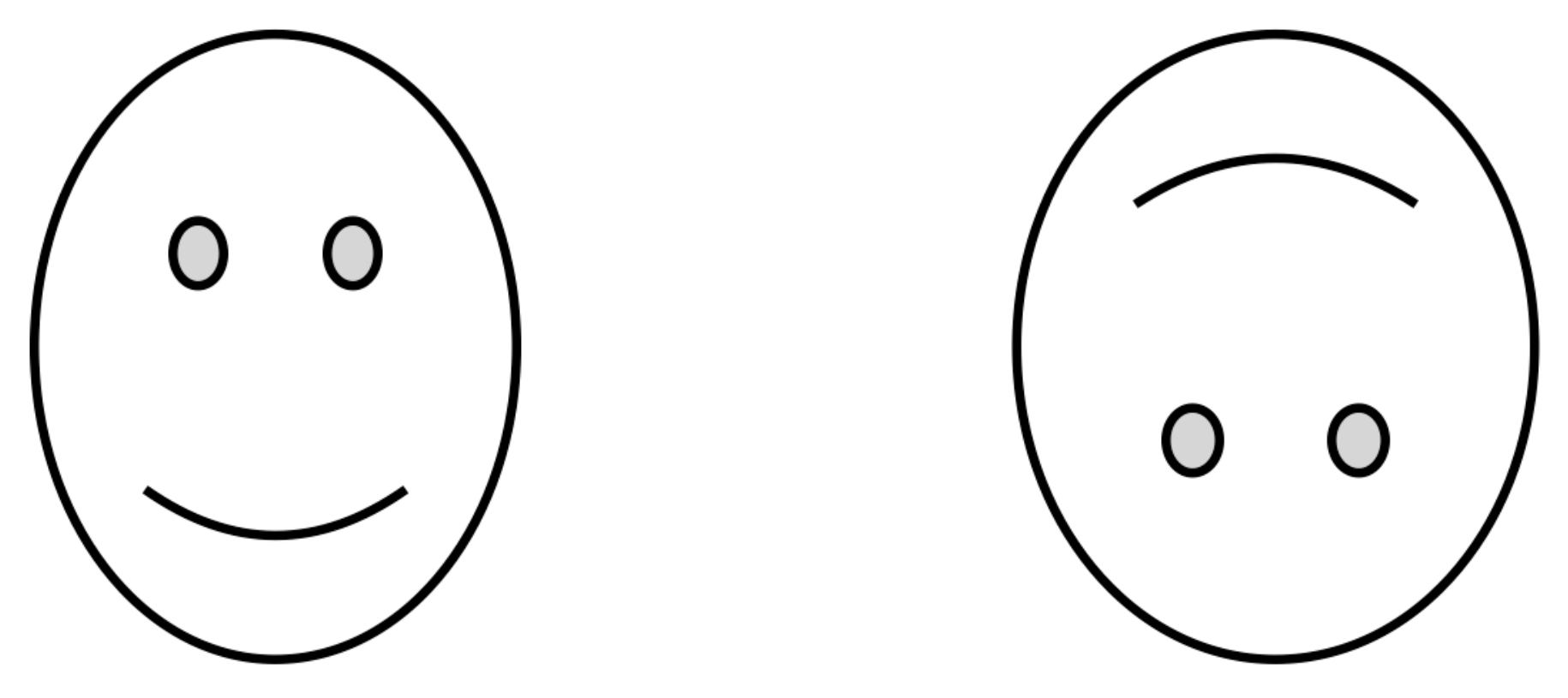} }
%
\caption{%
Schematic illustration of typical stimulus shown to observer on a single trial.
The observer has to choose the face that looks wider. The stimulus used in the experiment was a picture of the actor James Corden's face, with all background details removed (the illusion can be seen at http://illusionoftheyear.com/2010/the-fat-face-thin-fft-illusion). 
%
}
\label{figfaces}
\end{center}
\end{figure}
\fi

\section{Results} \label{secres}
\subsection*{Demonstration Experiment: Fat-Face Thin}
%
The CEPRD model described above was used to estimate the parameters $\theta$ for a simple demonstration experiment. On each trial, the observer was presented with a coloured picture of an upright face and an inverted face (see Figure \ref{figfaces}) on a computer screen, and was required to indicate which face appeared to be wider by pressing a left/right computer key. The faces remained visible until a response was made, and there was an interval of 0.5s between trials. For half of the trials, the reference stimulus was an upright face, and the comparison stimulus was an inverted version of the same face, and these were swapped for the other half of the trials. The width of the comparison image was determined by one of 21 stretch factors $s = 0.90, 0.91, . . . , 1.10$ (the heights of both stimuli were the same, and constant throughout the experiment). The stimulus strength was defined to be $x = s - 1$, so that $x$ varied between -0.1 and 0.1. For a given value of $s_{i}$, the observer was presented with the same stimulus pair for a total of $\ntrials = 20 $ trials. Stimuli were shown in random order, and the left/right position of reference/comparison stimuli was counterbalanced across trials.

For completeness, the mean parameter values are
$A = 0.848$, 
$K =  31.3$,
$x_{PSE} =   1.00$, 
which are consistent with values reported for other experiments using this type of model\cite{Shadlen2005}. 
The average reaction time across all observers is  
$\overline{\RTmean}=1.130$s, and the average decision time is 
$\overline{\RTmean}_{dec}=0.632$s. 

\ifFIG
\begin{figure}[b!] %
\begin{center}
{\includegraphics[width=0.99\textwidth] {\figurespath/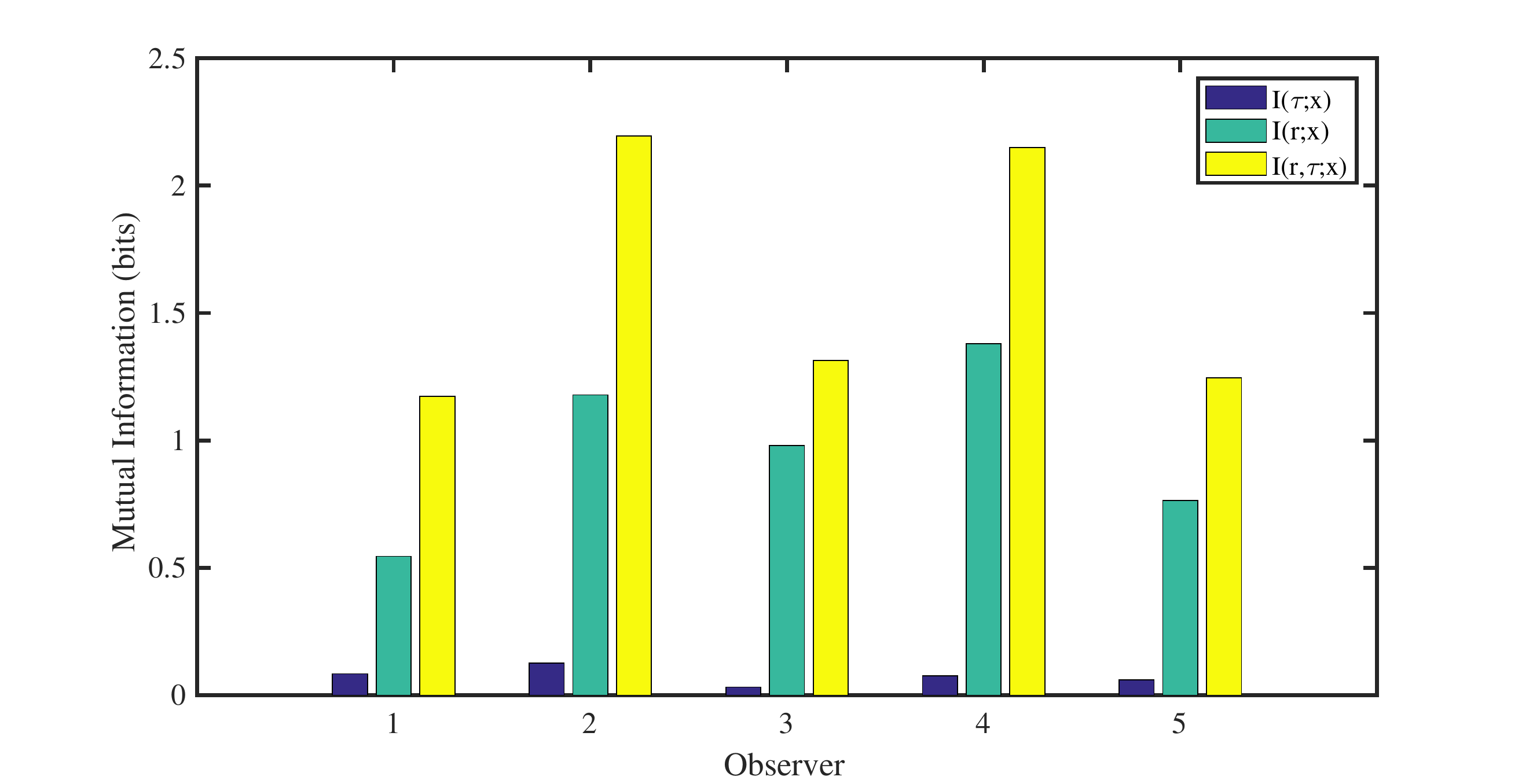} }%
\caption{
Mutual information between response per trial and stimulus strength for 5 observers, obtained (Equation \ref{eqSNRsingleRTy}). 
For each observer, reading from left to right, each bar represents  \\
(a, blue bar)   mutual information $I(\tau;x)$ between reaction time and stimulus strength $x$. \\
(b, green bar)    mutual information $I(r;x)$ between binary response and $x$. \\
(c, yellow bar)   total mutual information $I(\tau,r;x)$. 
}
\label{figMIall10observersA}
\end{center}
\end{figure}
\fi
        
The main results are summarised in Figure 
\ref{figMIall10observersA}.  The following estimates are means, where each mean is taken over all five observers, as shown in Figure \ref{figMIall10observersA}. 
The mean mutual information per trial between combined reaction time, binary response and stimulus strength is  
\bea
	I(y_{single};x) & = & 1.62 \text{ bits}. 
\eea
Combined with the mean decision time of 0.632 seconds, this implies the average observer acquires information at the rate of 
\bea
R = \frac{I(y_{single};x)}{\overline{\RTmean}_{dec}} = \frac{1.62}{0.632} =    2.56 \text{ bits/s}.
\eea    
The mean mutual information between  reaction time per trial and stimulus strength is  
\bea
	I(\tau;x) & = & 0.075 \text{ bits}. 
\eea
The mean mutual information between binary responses and  stimulus strength  is 
\bea
	I(r;x) & = & 0.970  \text{ bits}. 
\eea
If  reaction time and binary responses represented independent sources then the total mutual information would be 
\bea
	I_{sum} = I(\tau;x) + I(r;x) = .075+0.970 = 1.045 \text{ bits}.
\eea 
The combination of reaction times and binary responses provides more information
($I(y_{single};x) = 1.62$ bits)  about $x$ than the sum of their individual contributions ($I_{sum} =1.045$ bits). This, in turn, suggests that there is synergy between reaction time and binary responses. Even if this were not true, the finding that reaction times provide additional information means that making use of both reaction times and binary responses should increase the precision of parameter estimates, or (equivalently) substantially reduce the number experimental trials required. 

\subsection*{Shannon Competence and D-prime}
%
For a signal to noise ratio $SNR$, mutual $I$ information obeys the relation \cite{ShannonWeaverBook}
\bea
	I & \leq & 1/2 \jlog (1+ SNR) \text{ bits}, \label{eqsnra}
\eea
with equality if distributions are Gaussian, as is assumed here. The signal to noise ratio $SNR=S/N$, where $S$ and $N$ are standard symbols for signal and noise variance, respectively. 

To relate mutual information to more conventional measures of performance such as the discriminability measure $d'$ (d-prime), we note that  $d'^{2}=S/N$ \cite{BialekBook1996}, so that  
\bea
	I(r;x) & = & 1/2 \jlog (1+ d'^{2}) \text{ bits}. \label{eqsnr}
\eea
 Solving Equation \ref{eqsnr} for $d'$ yields 
\bea
	d' & = & (2^{2 I(r;x)} - 1)^{1/2}.
\eea
Given that $I(r;x)=0.970$, it follows that, if only binary responses were used to measure performance then this would have yielded a value of $d'=1.68$.

However, the mutual information $ I(y_{single};x)$ 
 takes account of both $RT$ and $P$, which 
allows a more general measure of performance to be defined by making use of the relation,
\bea
	I(y;x) & = & 1/2 \jlog (1+ s'^{2}) \text{ bits}. \label{eqsnrs}
\eea
Solving for $s'$ allows a new measure,  
{\em Shannon competence}, to be defined as
 \bea
	s' & = & (2^{2 I(y_{single};x)} - 1)^{1/2}.
\eea
The value of $I(y_{single};x)=1.62$ bits implies a Shannon competence of $s'=2.91$. Comparing this to $d'=1.68$ suggests that taking account of both $RT$ and $P$ almost doubles the estimated discriminability of stimuli in this experiment. 

\newpage
\section{Discussion}
%






Early attempts to incorporate both reaction time and binary responses into measures of performance tended to be {\em ad hoc} (see \cite{vandierendonck2017comparison} and  \cite{stafford2020quantifying}  for reviews). For example, the inverse efficiency score (IES) \cite{townsend1978methods} is the ratio IES=$RT/$(percent correct), which has not been found not to be justified \cite{bruyer2011combining}. Specifically, the sub-title of the paper \cite{bruyer2011combining} asks the question:  Is the inverse efficiency score (IES) a better dependent variable than the mean reaction time (RT) and the percentage of errors (PE)? According to that paper,  the answer seems to be $no$.  

Even though more recent attempts are sophisticated in many respects, they ignore the covariance between RT and binary responses (e.g. Equation 17 in \cite{bogacz2006physics},  and Equation 14 in \cite{bogacz2010}). Similarly, PRD models explicitly ignore such interactions, as explained in the next paragraph.
 
In summary, this paper presents four innovations.  
\begin{enumerate}
\item The covariant extended proportional rate diffusion (CEPRD) model 
was fitted by taking into account the full covariance structure of reaction times and binary responses. This contrasts with both Palmer et al's PRD model \cite{Shadlen2005}  and Stone's EPRD model \cite{StoneRTInfoTheory2014}, where model parameter values were estimated by minimising the product of reaction time variance and binary response variance. Thus, both of these previous models implicitly assume that the covariance between reaction times and binary responses is zero, an assumption which is  unwarranted from the results reported here. 
\item On a related theme, this represents an improvement on Stone's EPRD model  \cite{StoneRTInfoTheory2014}, which used independent estimates of information based either on $RT$ or binary responses (but not both) to estimate lower and upper bounds of mutual information $I_{single}$.   
\item By estimating the entropy of the joint distribution of reaction times and binary responses and comparing this to the residual entropy of the joint distribution after responses have been fitted to a model, we were able to estimate the mutual information between responses and stimulus strength. When divided by observer decision time, the information rate in bits/s was estimated.
\item This mutual information allows a particular observer-specific signal-to-noise ratio to be calculated, which can be used to estimate $s'$, a measure of discriminability that is exactly analogous to the conventional discriminability measure $d'$.
\end{enumerate}

Finally, note that the general strategy outlined in Section 2 does not depend on any particular model (e.g. EPRD). For example, model-free estimates the unconditional entropy $H(U)$ and of the conditional entropy $H(V)$ could be obtained from 
of a three-dimensional table, in which the axes are reaction times, binary responses and stimulus strength. 


 
 \newpage
 
 \appendix
 
\section{Appendix}
\subsection{The Shannon Information of a Single Response} \label{theappendix}
%
We have derived expressions for the Shannon information implicit in the average reaction time $RT_{i}$ and also in the average binary response, which is summarised as the proportion $P_{i}$ of comparison responses, for a stimulus strength $x_{i}$. 
Here, we derive an expression for the Shannon information associated with a single trial; first for reaction time, and then for binary responses.

As the number of trials at each stimulus strength is increased, so the variance in each mean reaction time  $RT$ decreases, and the central limit  theorem (CLM) ensures that the distribution of errors becomes increasingly Gaussian. The mutual information between two 
variables (e.g.  reaction time  and stimulus strength) depends on the signal to noise ratio $SNR$ 
%
\bea
	I & \leq & {1}/{2} \: \jlog  (1+SNR), \label{eqSNRnoneq}
\eea
where $SNR$ is the signal variance expressed as a fraction of the noise variance in the measurement \cite{ShannonWeaverBook}.

Invoking the CLM, we assume that the distribution of differences $\Delta \tau$ between individual reaction times $\tau$ and the mean reaction time   (at one stimulus strength) is Gaussian. Because the mutual information is defined as 
 the  entropy of $\tau$ minus the  entropy of the noise $\Delta \tau$ in $\tau$, we can assume equality in Equation \ref{eqSNRnoneq} \cite{BialekBook1996}. In fact, we do not need to rely on the central limit theorem here, because even if  the perturbing noise is not Gaussian, Shannon's Theorem 18 \cite{ShannonWeaverBook}
 implies equality  in Equation \ref{eqSNRnoneq}, so that
\bea
	I & = & {1}/{2} \: \jlog  (1+SNR) \bits. \label{eqSNR}
\eea

We already have a value for the mutual information $I({\RTmean} ; x)$, so we can re-arrange Equation \ref{eqSNR} to find the SNR associated with $\RTmean$ 
\bea
	SNR_{\RTmean} & = &  2^{2I({x,\RTmean})}-1 \bits. \label{eqSNRmean}
\eea
However, the mutual information $I({x,\RTmean})$  obtained tells us how much average Shannon information each {\em mean} reaction time provides  about stimulus strength, whereas we want to know how much average  information each {\em individual} reaction time  provides  about stimulus strength. Because the value of SNR in Equation \ref{eqSNRmean} is based on mean RTs, each of which involves $N_{i}$ trials, the variance of the measurement noise has been reduced by a factor of $N_{i}$ relative to the noise in the reaction time  of a single trial (provided this noise is iid). This implies that the value of SNR for a single trial is
\bea
	SNR_{\tau} 	& = & SNR_{\RTmean} /N_{i}\\
				& = & (2^{2 I(\RTmean;x) }-1)/N_{i} \bits.
\eea
If we substitute $SNR_{\tau}$ into Equation \ref{eqSNR} then we obtain an estimate of the average Shannon information $I({x,\tau})$ implicit in the observer's reaction time in a single trial
\bea
	I({\tau;x}) & = & \frac{1}{2} \jlog \left[ 1+ \frac{(2^{2 I({\RTmean;x})}-1)}{N_{i}} \right] \bits. \label{eqSNRsingleRT}
\eea
By analogy, the average Shannon information $I({x,r})$ implicit in the observer's binary response $r$ in a single trial is 
\bea
	I(r;x) & = & \frac{1}{2} \jlog \left[ 1+\frac{(2^{2 I({P;x})}-1)}{N_{i}} \right] \bits. \label{eqSNRsingleP}
\eea 

\fi


\end{document}